\documentclass[preprint,prd,aps,showpacs,showkeys,nofootinbib]{revtex4}
\usepackage{graphicx}
\usepackage{dcolumn}
\usepackage{bm}
\begin{document}
\title{The one loop corrections to the neutrino masses in BLMSSM}
\author{Shu-Min Zhao$^1$\footnote{zhaosm@hbu.edu.cn}, Tai-Fu Feng$^{1}$\footnote{fengtf@hbu.edu.cn}, Xing-Xing Dong$^1$,
 Hai-Bin Zhang$^{1}$, Guo-Zhu Ning$^1$, Tao-Guo$^2$}
\affiliation{$^1$ Department of Physics, Hebei University, Baoding 071002,
China\\
$^2$ School of Mathematics and Science, Hebei University of Geosciences, Shijiazhuang 050031, China}
\date{\today}
\begin{abstract}
The neutrino masses and mixings are studied in the model
which is the supersymmetric extension of the
standard model with local gauged baryon
and lepton numbers(BLMSSM).
At tree level the neutrinos can obtain tiny masses through the See-Saw mechanism
in the BLMSSM.
The one-loop corrections to the neutrino masses and mixings are important, and they are studied
in this work with the mass insertion approximation.
We study the numerical results and discuss the allowed parameter space of BLMSSM.
It can contribute to study the neutrino masses and to explore the new physics
beyond the standard model(SM).

\end{abstract}

\pacs{13.40.Em, 12.60.-i}
\keywords{BLMSSM, neutrino mass, mass insertion}

\maketitle

\section{Introduction}
\indent\indent
When the CP-even Higgs $h^0$ with $m_{h^0}=125.7{\rm GeV}$ was detected by LHC in 2012\cite{Higgs},
all the particles in SM have been founded.
Though SM obtains large successes, it is unable to explain some phenomena.
For example, SM can not explain the neutrino masses and
their mixing pattern\cite{NeuExp}, because in SM there are only three left-handed neutrinos with zero mass. The anomalous neutrino data from
both solar and atmospheric neutrino experiments promote the study of neutrino masses and lepton flavor violating processes.
The authors give out the global analyses of neutrino oscillation experiments in their work, and
the present $3\sigma$ limits for the neutrino experiment data are\cite{15NGLF}
\begin{eqnarray}
&&0.0188\leq\sin^2\theta_{13}\leq0.0251,\nonumber\\&&0.270\leq\sin^2\theta_{12}\leq0.344,\nonumber\\&&
0.385\leq\sin^2\theta_{23}\leq0.644,\nonumber\\
&&7.02\times 10^{-5} {\rm eV}^2\leq\Delta m_{\odot}^2 \leq 8.09\times 10^{-5} {\rm eV}^2,\nonumber\\&&
2.325\times 10^{-3} {\rm eV}^2\leq\Delta m_{A}^2(\texttt{NO}) \leq2.599\times 10^{-3} {\rm eV}^2,\nonumber\\&&
-2.590\times 10^{-3} {\rm eV}^2\leq\Delta m_{A}^2(\texttt{IO}) \leq-2.307\times 10^{-3} {\rm eV}^2.
\label{neu-oscillations2}
\end{eqnarray}
For the mixing pattern, there are two large mixing angles and one small mixing angle.  To explain the results in Eq.(\ref{neu-oscillations2}), a theory beyond the SM is necessary.
Therefore, the neutrino sector is a natural testing ground for the new models beyond the SM.

For the new physics, the supersymmetric extension of the SM is a popular choice. The discrete symmetry known as $R$-parity is
defined as $R_p=(-1)^{L+3B+2S}$ with $L(B),S$ denoting the lepton (baryon) number and the spin of the particle\cite{RP}.
The minimal supersymmetric extension of the SM (MSSM)\cite{MSSM} has been studied for many years by
theoretical physicists. MSSM with $R$-parity conservation has some short comings,
where neither $\mu$ problem nor the observed neutrino masses
can be explained. $R$-parity violation can be obtained from $L$ breaking, $B$ breaking, both $L$ and $B$ breaking.
 In general, to explain the neutrino experiment data, the lepton number should be broken.

In $R$-parity violating supersymmetric models with generic soft supersymmetry
breaking terms\cite{RPSUSY}, neutrinos and neutralinos mix together at tree
level. Therefore, one neutrino gets small mass through the see-saw mechanism\cite{seesaw}.
The loop diagrams including lepton number violating effects provide masses to the other neutrinos.
Taking into account the one loop effects, the authors research the neutrino masses and mixings.
In $\mu\nu SSM$ three right-handed neutrino superfields are introduced\cite{munuSSM1,munuSSM2}, and it
can solve the $\mu$ problem. In this model\cite{munuSSMOL},  the neutrino masses and mixings are studied at one loop level,
and significative results are obtained.

The BLMSSM is the minimal supersymmetric extension of
the SM with local gauged $B$ and $L$, which is spontaneously broken at TeV scale\cite{BLMSSM1}.
This model was first proposed by the authors in Ref.\cite{BLMSSM0}.
So BLMSSM is $R$-parity
 violating and can explain the asymmetry of matter-antimatter in the universe.
 In BLMSSM, the authors study the lightest CP-even Higgs mass and  the
 decays  $h^0\rightarrow\gamma\gamma$, $h^0\rightarrow ZZ (WW)$\cite{BLMSSM2}. The one loop and two loop Barr-Zee type contributions
 to muon MDM and charged lepton flavor violating processes are also discussed\cite{BLBarZee}.
 Considering the CP-violation,
 we study neutron EDM, lepton EDM and $B^0-\bar{B}^0$ mixing in this model\cite{BLMSSM3}.

 In the BLMSSM, because of the introduced
 right-handed neutrino fields, the three light neutrinos obtain tiny masses at tree level through the
 see-saw mechanism, which is shown in our previous work\cite{BLCB}. Here, using the mass insertion approximation we consider one loop corrections to the
 neutrino mass mixing matrix. The one loop corrections are important, especially for the light neutrinos.

After this introduction, in Section 2 we briefly introduce the BLMSSM. With mass insertion method, the one loop corrections
 to neutrino mass matrix are shown in Section 3. Section 4 is devoted to the numerical analysis.
 The summary is given out in Section 5.

\section{some content of BLMSSM}
The gauge symmetry of BLMSSM is $SU(3)_{C}\otimes SU(2)_{L}\otimes U(1)_{Y}\otimes U(1)_{B}\otimes U(1)_{L}$.
Compared with MSSM, BLMSSM includes many new fields\cite{BLMSSM1}:

1. the exotic quarks $(\hat{Q}_{4},\hat{U}_{4}^c,\hat{D}_{4}^c,\hat{Q}_{5}^c,\hat{U}_{5},\hat{D}_{5})$ used to cancel $B$ anomaly,

2. the exotic leptons $(\hat{L}_{4},\hat{E}_{4}^c,\hat{N}_{4}^c,\hat{L}_{5}^c,\hat{E}_{5},\hat{N}_{5})$ used to cancel $L$ anomaly,

3. the exotic Higgs $\hat{\Phi}_{L},\hat{\varphi}_{L}$ introduced to break $L$ spontaneously with nonzero vacuum expectation values (VEVs),

4. the exotic Higgs $\hat{\Phi}_{B},\hat{\varphi}_{B}$ introduced to break $B$ spontaneously with nonzero VEVs,

5. the superfields $\hat{X}$ and $\hat{X}^\prime$ used to make the exotic quarks unstable,

6. the right-handed neutrinos $N_R^c$ introduced to provide tiny masses of neutrinos through see-saw mechanism.

The lightest mass eigenstate of the mixed mass matrix for $\hat{X}$ and $\hat{X}^\prime$ could be a dark matter candidate.
These new fields are shown particularly in the table1.
 \begin{table}[htdp]
\caption{ The new fields in the BLMSSM: the exotic quarks, exotic leptons, exotic Higgs, $X$ superfields and right-handed neutrinos.}
\begin{center}
\begin{tabular}{|c|c|c|c|c|c|}
\hline
Superfields & $SU(3)_C$ & $SU(2)_L$ & $U(1)_Y$ & $U(1)_B$ & $U(1)_L$\\
\hline
\hline
$\hat{Q}_4$ & 3 & 2 & 1/6 & $B_4$ & 0 \\
\hline
$\hat{U}^c_4$ & $\bar{3}$ & 1 & -2/3 & -$B_4$ & 0 \\
\hline
$\hat{D}^c_4$ & $\bar{3}$ & 1 & 1/3 & -$B_4$ & 0 \\
\hline
$\hat{Q}_5^c$ & $\bar{3}$ & 2 & -1/6 & -$(1+B_4)$ & 0 \\
\hline
$\hat{U}_5$ & $3$ & 1 & 2/3 &  $1 + B_4$ & 0 \\
\hline
$\hat{D}_5$ & $3$ & 1 & -1/3 & $1 + B_4$ & 0 \\
\hline
$\hat{L}_4$ & 1 & 2 & -1/2 & 0 & $L_4$ \\
\hline
$\hat{E}^c_4$ & 1 & 1 & 1 & 0 & -$L_4$ \\
\hline
$\hat{N}^c_4$ & 1 & 1 & 0 & 0 & -$L_4$ \\
\hline
$\hat{L}_5^c$ & 1 & 2 & 1/2 & 0 & -$(3 + L_4)$ \\
\hline
$\hat{E}_5$ & 1 & 1 & -1 & 0 & $3 + L_4$ \\
\hline
$\hat{N}_5$ & 1 & 1 & 0 & 0 & $3 + L_4$ \\
\hline
$\hat{\Phi}_B$ & 1 & 1 & 0 & 1 & 0 \\
\hline
$\hat{\varphi}_B$ & 1 & 1 & 0 & -1 & 0 \\
\hline
$\hat{\Phi}_L$ & 1 & 1 & 0 & 0 & -2 \\
\hline
$\hat{\varphi}_L$ & 1 & 1 & 0 & 0 & 2 \\
\hline
$\hat{X}$ & 1 & 1 & 0 & $2/3 + B_4$ & 0 \\
\hline
$\hat{X'}$ & 1 & 1 & 0 & $-(2/3 + B_4)$ & 0 \\
\hline
$\hat{N}_R^c$ & 1 & 1 & 0 & 0 & -1 \\
\hline
\end{tabular}
\end{center}
\label{quarks}
\end{table}

 The superpotential of BLMSSM is written as \cite{BLMSSM1,BLMSSM2}
\begin{eqnarray}
&&{\cal W}_{{BLMSSM}}={\cal W}_{{MSSM}}+{\cal W}_{B}+{\cal W}_{L}+{\cal W}_{X}\;,
\label{superpotential1}
\end{eqnarray}
with ${\cal W}_{{MSSM}}$ denoting the superpotential of the MSSM. The concrete forms of ${\cal W}_{B}, {\cal W}_{L}$ and $ {\cal W}_{X}$
are
\begin{eqnarray}
&&{\cal W}_{B}=\lambda_{Q}\hat{Q}_{4}\hat{Q}_{5}^c\hat{\Phi}_{B}+\lambda_{U}\hat{U}_{4}^c\hat{U}_{5}
\hat{\varphi}_{B}+\lambda_{D}\hat{D}_{4}^c\hat{D}_{5}\hat{\varphi}_{B}+\mu_{B}\hat{\Phi}_{B}\hat{\varphi}_{B}
\nonumber\\
&&\hspace{1.2cm}
+Y_{{u_4}}\hat{Q}_{4}\hat{H}_{u}\hat{U}_{4}^c+Y_{{d_4}}\hat{Q}_{4}\hat{H}_{d}\hat{D}_{4}^c
+Y_{{u_5}}\hat{Q}_{5}^c\hat{H}_{d}\hat{U}_{5}+Y_{{d_5}}\hat{Q}_{5}^c\hat{H}_{u}\hat{D}_{5}\;,
\nonumber\\
&&{\cal W}_{L}=Y_{{e_4}}\hat{L}_{4}\hat{H}_{d}\hat{E}_{4}^c+Y_{{\nu_4}}\hat{L}_{4}\hat{H}_{u}\hat{N}_{4}^c
+Y_{{e_5}}\hat{L}_{5}^c\hat{H}_{u}\hat{E}_{5}+Y_{{\nu_5}}\hat{L}_{5}^c\hat{H}_{d}\hat{N}_{5}
\nonumber\\
&&\hspace{1.2cm}
+Y_{\nu}\hat{L}\hat{H}_{u}\hat{N}^c+\lambda_{{N^c}}\hat{N}^c\hat{N}^c\hat{\varphi}_{L}
+\mu_{L}\hat{\Phi}_{L}\hat{\varphi}_{L}\;,
\nonumber\\
&&{\cal W}_{X}=\lambda_1\hat{Q}\hat{Q}_{5}^c\hat{X}+\lambda_2\hat{U}^c\hat{U}_{5}\hat{X}^\prime
+\lambda_3\hat{D}^c\hat{D}_{5}\hat{X}^\prime+\mu_{X}\hat{X}\hat{X}^\prime\;.
\label{superpotential-BL}
\end{eqnarray}

The soft breaking terms $\mathcal{L}_{{soft}}$ of the BLMSSM can be found in the works\cite{BLMSSM0,BLMSSM1, BLMSSM2}.
\begin{eqnarray}
&&{\cal L}_{{soft}}={\cal L}_{{soft}}^{MSSM}-(m_{{\tilde{\nu}^c}}^2)_{{IJ}}\tilde{N}_I^{c*}\tilde{N}_J^c
-m_{{\tilde{Q}_4}}^2\tilde{Q}_{4}^\dagger\tilde{Q}_{4}-m_{{\tilde{U}_4}}^2\tilde{U}_{4}^{c*}\tilde{U}_{4}^c
-m_{{\tilde{D}_4}}^2\tilde{D}_{4}^{c*}\tilde{D}_{4}^c
\nonumber\\
&&\hspace{1.3cm}
-m_{{\tilde{Q}_5}}^2\tilde{Q}_{5}^{c\dagger}\tilde{Q}_{5}^c-m_{{\tilde{U}_5}}^2\tilde{U}_{5}^*\tilde{U}_{5}
-m_{{\tilde{D}_5}}^2\tilde{D}_{5}^*\tilde{D}_{5}-m_{{\tilde{L}_4}}^2\tilde{L}_{4}^\dagger\tilde{L}_{4}
-m_{{\tilde{\nu}_4}}^2\tilde{N}_{4}^{c*}\tilde{N}_{4}^c
\nonumber\\
&&\hspace{1.3cm}
-m_{{\tilde{e}_4}}^2\tilde{E}_{_4}^{c*}\tilde{E}_{4}^c-m_{{\tilde{L}_5}}^2\tilde{L}_{5}^{c\dagger}\tilde{L}_{5}^c
-m_{{\tilde{\nu}_5}}^2\tilde{N}_{5}^*\tilde{N}_{5}-m_{{\tilde{e}_5}}^2\tilde{E}_{5}^*\tilde{E}_{5}
-m_{{\Phi_{B}}}^2\Phi_{B}^*\Phi_{B}
\nonumber\\
&&\hspace{1.3cm}
-m_{{\varphi_{B}}}^2\varphi_{B}^*\varphi_{B}-m_{{\Phi_{L}}}^2\Phi_{L}^*\Phi_{L}
-m_{{\varphi_{L}}}^2\varphi_{L}^*\varphi_{L}-\Big(m_{B}\lambda_{B}\lambda_{B}
+m_{L}\lambda_{L}\lambda_{L}+h.c.\Big)
\nonumber\\
&&\hspace{1.3cm}
+\Big\{A_{{u_4}}Y_{{u_4}}\tilde{Q}_{4}H_{u}\tilde{U}_{4}^c+A_{{d_4}}Y_{{d_4}}\tilde{Q}_{4}H_{d}\tilde{D}_{4}^c
+A_{{u_5}}Y_{{u_5}}\tilde{Q}_{5}^cH_{d}\tilde{U}_{5}+A_{{d_5}}Y_{{d_5}}\tilde{Q}_{5}^cH_{u}\tilde{D}_{5}
\nonumber\\
&&\hspace{1.3cm}
+A_{{BQ}}\lambda_{Q}\tilde{Q}_{4}\tilde{Q}_{5}^c\Phi_{B}+A_{{BU}}\lambda_{U}\tilde{U}_{4}^c\tilde{U}_{5}\varphi_{B}
+A_{{BD}}\lambda_{D}\tilde{D}_{4}^c\tilde{D}_{5}\varphi_{B}+B_{B}\mu_{B}\Phi_{B}\varphi_{B}
+h.c.\Big\}
\nonumber\\
&&\hspace{1.3cm}
+\Big\{A_{{e_4}}Y_{{e_4}}\tilde{L}_{4}H_{d}\tilde{E}_{4}^c+A_{{\nu_4}}Y_{{\nu_4}}\tilde{L}_{4}H_{u}\tilde{N}_{4}^c
+A_{{e_5}}Y_{{e_5}}\tilde{L}_{5}^cH_{u}\tilde{E}_{5}+A_{{\nu_5}}Y_{{\nu_5}}\tilde{L}_{5}^cH_{d}\tilde{N}_{5}
\nonumber\\
&&\hspace{1.3cm}
+A_{N}Y_{\nu}\tilde{L}H_{u}\tilde{N}^c+A_{{N^c}}\lambda_{{N^c}}\tilde{N}^c\tilde{N}^c\varphi_{L}
+B_{L}\mu_{L}\Phi_{L}\varphi_{L}+h.c.\Big\}
\nonumber\\
&&\hspace{1.3cm}
+\Big\{A_1\lambda_1\tilde{Q}\tilde{Q}_{5}^cX+A_2\lambda_2\tilde{U}^c\tilde{U}_{5}X^\prime
+A_3\lambda_3\tilde{D}^c\tilde{D}_{5}X^\prime+B_{X}\mu_{X}XX^\prime+h.c.\Big\}\;,
\label{soft-breaking}
\end{eqnarray}

When the Higgs fields including $SU(2)_L$ doublets ($H_{u}$ $H_{d}$) and  $SU(2)_L$ singlets
($\Phi_{L}$, $\varphi_{L}$, $\Phi_{B}$, $\varphi_{B}$)
 obtain nonzero VEVs, the local gauge symmetry $SU(2)_{L}\otimes U(1)_{Y}\otimes U(1)_{B}\otimes U(1)_{L}$
  breaks down to the electromagnetic symmetry $U(1)_{e}$.
The $SU(2)_L$ doublets $H_{u},\;H_{d}$ are defined as
\begin{eqnarray}
&&H_{u}=\left(\begin{array}{c}H_{u}^+\\{1\over\sqrt{2}}\Big(\upsilon_{u}+H_{u}^0+iP_{u}^0\Big)\end{array}\right)\;,~~~~
H_{d}=\left(\begin{array}{c}{1\over\sqrt{2}}\Big(\upsilon_{d}+H_{d}^0+iP_{d}^0\Big)\\H_{d}^-\end{array}\right)\;,
\end{eqnarray}
with nonzero VEVs $\upsilon_{u},\;\upsilon_{d}$.
The $SU(2)_L$ singlets $(\Phi_{B},\varphi_{B},\Phi_{L},
\varphi_{L})$ have nonzero VEVs
 $\upsilon_{{B}},\overline{\upsilon}_{{B}}, \upsilon_{L},\overline{\upsilon}_{L}$.
  \begin{eqnarray}
&&\Phi_{B}={1\over\sqrt{2}}\Big(\upsilon_{B}+\Phi_{B}^0+iP_{B}^0\Big)\;,~~~~~~~~~
\varphi_{B}={1\over\sqrt{2}}\Big(\overline{\upsilon}_{B}+\varphi_{B}^0+i\overline{P}_{B}^0\Big)\;,
\nonumber\\
&&\Phi_{L}={1\over\sqrt{2}}\Big(\upsilon_{L}+\Phi_{L}^0+iP_{L}^0\Big)\;,~~~~~~~~~~
\varphi_{L}={1\over\sqrt{2}}\Big(\overline{\upsilon}_{L}+\varphi_{L}^0+i\overline{P}_{L}^0\Big)\;.
\label{VEVs}
\end{eqnarray}
\section{the coupling}
Because in the BLMSSM neutrinos are Majorana particles, we can use the following expressions for the neutrinos.
In the base $( \psi_ {\nu^I_L},\psi_{N^{cI}_R} )$, the formulae for mass mixing matrix and mass eigenstates
are shown here\cite{BLMSSM3}.

\begin{eqnarray}
&&Z_{N_\nu}^{T}\left(\begin{array}{cc}
  0&\frac{v_u}{\sqrt{2}}(Y_{\nu})^{IJ} \\
   \frac{v_u}{\sqrt{2}}(Y^{T}_{\nu})^{IJ}  & \frac{\bar{v}_L}{\sqrt{2}}(\lambda_{N^c})^{IJ}
    \end{array}\right) Z_{N_\nu}=diag(m_{\nu^\alpha}), ~~ \alpha=1\dots 6,~~ I,J=1,2,3,
    \nonumber\\&& \psi_{\nu^I_L}=Z_{N_\nu}^{I\alpha}k_{N_\alpha}^0,~~~~
   \psi_{N^{cI}_R}=Z_{N_\nu}^{(I+3)\alpha}k_{N_\alpha}^0,~~~~
  \chi_{N_\alpha}^0= \left(\begin{array}{c}
 k_{N_\alpha}^0\\ \bar{k}_{N_\alpha}^0
    \end{array}\right). \label{massnu}
\end{eqnarray}
 $\chi_{N_\alpha}^0 (\alpha=1\dots 6)$ denote the mass eigenstates of the neutrino fields mixed by left-handed and right-handed neutrinos.

The exotic gauginos ($\lambda_L$, $\lambda_B$) and exotic Higgs super fields ($\psi_{\Phi_L},\psi_{\varphi_L}$, $\psi_{\Phi_B},\psi_{\varphi_B}$)
are introduced in BLMSSM. They mix together leading to six new neutralinos beyond MSSM.
However, the six new neutralinos do not mix with the four MSSM neutralinos.
$\lambda_L$
(the superpartners of the new lepton boson) and $\psi_{\Phi_L},\psi_{\varphi_L}$
 (the superpartners of the $SU(2)_L$ singlets $\Phi_L,\varphi_L$) mix and they produce three lepton neutralinos.
 In the basis $(i\lambda_L,\psi_{\Phi_L},\psi_{\varphi_L})$, the mass mixing matrix of lepton neutralinos is\cite{BLMSSM3}
 \begin{equation}
\left(     \begin{array}{ccc}
  2M_L &2v_Lg_L &-2\bar{v}_Lg_L\\
   2v_Lg_L & 0 &-\mu_L\\-2\bar{v}_Lg_L&-\mu_L &0
    \end{array}\right).\label{LN}
   \end{equation}
   To get the mass eigenstates for lepton neutralinos, we use the rotation matrix $Z_{N_L}$ to diagonalize the mass mixing matrix in Eq.(\ref{LN}).

Similarly, three baryon neutralinos  are produced from $\lambda_B$
(the superpartners of the new baryon boson) and $\psi_{\Phi_B},\psi_{\varphi_B}$
 (the superpartners of the $SU(2)_L$ singlets $\Phi_B,\varphi_B$).
We show the mass mixing matrix of baryon neutralinos here in the basis $(i\lambda_B,\psi_{\Phi_B},\psi_{\varphi_B})$
 \begin{equation}
\left(     \begin{array}{ccc}
  2M_B &2v_Bg_B &-2\bar{v}_Bg_B\\
   2v_Bg_B & 0 &-\mu_B\\-2\bar{v}_Bg_B&-\mu_B &0
    \end{array}\right).\label{LB}
   \end{equation}
To obtain three baryon neutrino masses, we use the rotation matrix
$Z_{N_B}$ to diagonalize the mass mixing matrix in Eq.(\ref{LB}).

From the supperpotential $\mathcal{W }_L$ and the interactions of gauge and matter multiplets
$ig\sqrt{2}T^a_{ij}(\lambda^a\psi_jA_i^*-\bar{\lambda}^a\bar{\psi}_iA_j)$, we obtain the couplings with
light neutrinos at tree level.
\begin{eqnarray}
&&\mathcal{L}_L(\nu)=-Y_l^{IJ}Z_-^{2i}Z_{N_\nu}^{I\alpha}\bar{\chi}_i^+\omega_-\chi^0_{N_\alpha}\tilde{e}_R^{JC}
-Y_\nu^{IJ}Z_N^{4i}Z_{N_\nu}^{I\alpha}\bar{\chi}_i^0\omega_-\chi^0_{N_\alpha}\tilde{N}^{cJ}_R
\nonumber\\&&\hspace{1.6cm}-Y_l^{IJ}Z_{N_\nu}^{I\alpha}\bar{e}^J\omega_-\chi^0_{N_\alpha}H_d^2
-Y_\nu^{IJ}Z_{N_\nu}^{I\alpha}Z_{N_\nu}^{(3+J)\beta}\bar{\chi}^0_{N_\beta}\omega_-\chi^0_{N_\alpha}H_u^2
\nonumber\\&&\hspace{1.6cm}
-g_2Z_-^{1i}Z_{N_\nu}^{I\alpha}\bar{\chi}_i^+\omega_-\chi^0_{N_\alpha}\tilde{e}^{-*}_L
  +\sqrt{2}g_LZ_{N_L}^{1i}Z_{N_\nu}^{I\alpha}\bar{\chi}^0_{Li}\omega_-\chi^0_{N_\alpha}\tilde{\nu}^*_L
  \nonumber\\&&\hspace{1.6cm}-\frac{1}{\sqrt{2}}Z_{N_\nu}^{I\alpha}\Big(g_2Z_N^{2i}-g_1Z_N^{1i}\Big)
  \bar{\chi}_i^0\omega_-\chi^0_{N_\alpha}\tilde{\nu}^{I*}_L+h.c. \label{LLnu}
\end{eqnarray}

In the same way, the couplings related to heavy neutrinos are also obtained
\begin{eqnarray}
&&\mathcal{L}_H(\nu)=Y_\nu^{IJ}Z_{N_\nu}^{(J+3)\alpha}\bar{\chi}^0_{N_\alpha}\omega_-e^IH_u^1
+Y_\nu^{IJ}Z_+^{2i}Z_{N_\nu}^{(J+3)\alpha}\bar{\chi}^0_{N_\alpha}\omega_-\chi_i^+\tilde{e}_L^I
\nonumber\\&&\hspace{1.6cm}-
Y_\nu^{IJ}Z_N^{4i}Z_{N_\nu}^{(J+3)\alpha}\bar{\chi}^0_{N_\alpha}\omega_-\chi_i^0\tilde{\nu}_L^I
-\lambda_{N^c}^{IJ}Z_{N_\nu}^{(I+3)\alpha}Z_{N_\nu}^{(J+3)\beta}\bar{\chi}^0_{N_\alpha}
\omega_-\chi^0_{N_\beta}\varphi_L\nonumber\\&&\hspace{1.6cm}
-\Big((\lambda_{N^c}^{IJ}+\lambda_{N^c}^{JI})Z_{N_L}^{3i}
+\sqrt{2}g_LZ_{N_L}^{1i}\delta^{IJ}\Big)Z_{N_\nu}^{(I+3)\alpha}
\bar{\chi}^0_{N_\alpha}\omega_-\chi^0_{L_i} N_R^{cJ}+h.c. \label{LHnu}
\end{eqnarray}

\section{the one loop corrections to neutrino mass matrix}
The neutrino Yukawa couplings $(Y_\nu)^{IJ},(I,J=1,2,3)$  are
much smaller than the other
couplings. For Eqs.(\ref{LLnu})(\ref{LHnu}), the terms $-Y_\nu^{IJ}Z_N^{4i}Z_{N_\nu}^{I\alpha}\bar{\chi}_i^0\omega_-\chi^0_{N_\alpha}\tilde{N}^{cJ}_R$
 $ -Y_\nu^{IJ}Z_{N_\nu}^{I\alpha}Z_{N_\nu}^{(3+J)\beta}\bar{\chi}^0_{N_\beta}\omega_-\chi^0_{N_\alpha}H_u^2$
in $\mathcal{L}_L(\nu)$, and $Y_\nu^{IJ}Z_{N_\nu}^{(J+3)\alpha}\bar{\chi}^0_{N_\alpha}\omega_-e^IH_u^1
+Y_\nu^{IJ}Z_+^{2i}Z_{N_\nu}^{(J+3)\alpha}\bar{\chi}^0_{N_\alpha}\omega_-\chi_i^+\tilde{e}_L^I
-Y_\nu^{IJ}Z_N^{4i}Z_{N_\nu}^{(J+3)\alpha}\bar{\chi}^0_{N_\alpha}\omega_-\chi_i^0\tilde{\nu}_L^I$ in $\mathcal{L}_H(\nu)$
can be neglected safely, because they are suppressed by $Y_\nu$ compared with the other terms.
The $Y_{\nu}$ in the neutrino mass mixing matrix at tree level is not neglected.  One can find that $Z_{N_\nu}$ is the function of
$Y_{\nu}$ from Eq.(\ref{massnu}). That is to say, $Z_{N_\nu}$ is relevant to the chiral symmetry breaking terms.

 Using the mass insertion approximation\cite{MIA},
 we deduce the neutrino mass corrections from the virtual slepton-chargino at one loop level
\begin{eqnarray}
&&\delta (m_\nu)_{\alpha\theta}(\tilde{e}_L,\tilde{e}_R,\chi_{i}^{\pm})=
\frac{m_{\chi_i^{\pm}}}{\Lambda^2}\Big(Z_{N_\nu}^{I\alpha}Z_{N_\nu}^{J\theta}Y_l^IY_l^J(Z_-^{2i})^2
\delta (M^2_{\tilde{L}})^{IJ}_{RR}I^0_{111}(x_{\chi_i^{\pm}},x_{\tilde{e}_R^K},x_{\tilde{e}_R^I})
\nonumber\\&&\hspace{3.6cm}+(Z_{N_\nu}^{I\theta}Z_{N_\nu}^{J\alpha}+Z_{N_\nu}^{I\alpha}Z_{N_\nu}^{J\theta})g_2Y_l^IZ_-^{1i}Z_-^{2i}\delta (M^2_{\tilde{L}})^{IJ}_{RL}I^0_{111}(x_{\chi_{i}^{\pm}},x_{\tilde{e}_R^I},x_{\tilde{e}_L^J})\nonumber\\&&\hspace{3.6cm}
+Z_{N_\nu}^{I\alpha}Z_{N_\nu}^{J\theta}g_2^2(Z_-^{1i})^2\delta (M^2_{\tilde{L}})^{IJ}_{LL}I^0_{111}(x_{\chi_i^{\pm}},x_{\tilde{e}_L^I},x_{\tilde{e}_L^J})
\Big).\label{SLchar}
\end{eqnarray}
The one loop function $I_{111}^{0}(x_1,x_2,x_3)$ is defined from the following formula
\begin{eqnarray}
&&i\int\frac{dk^4}{(2\pi)^4}\frac{1}{k^2-m_1^2}\frac{1}{k^2-m_2^2}\frac{1}{k^2-m_3^2}=\frac{1}{\Lambda^2}I^0_{111}(x_1,x_2,x_3),
\end{eqnarray}
with $\Lambda$ representing the energy scale of the new physics and $x_i=\frac{m_i^2}{\Lambda^2}$ for $i=1,2,3$.
In Eq.(12), it seems that the results have nothing to do with the chiral symmetry breaking terms.
In fact, Eq.(12) includes $Z_{N_\nu}$ which is the function of nonzero $Y_\nu$. Therefore, Eq.(12) includes
the chiral symmetry breaking terms and gives corrections to the neutrino mass mixing matrix.

In the same way, the neutrino mass corrections from the virtual sneutrino-lepton neutralino and sneutrino-neutralino are obtained
\begin{eqnarray}
&&\delta (m_\nu)_{\alpha\theta}(\tilde{\nu}_L,\tilde{N}^c_R,\chi_{L_i}^{0})=
\frac{m_{\chi_{L_i}^{0}}}{\Lambda^2}\Big(2g_L^2
(Z_{N_L}^{1i})^2Z_{N_\nu}^{I\alpha}Z_{N_\nu}^{J\theta}
\delta (M^2_{\tilde{\nu}})^{IJ}_{LL}I^0_{111}(x_{\chi_{L_i}^{0}},x_{\tilde{\nu}_L^I},x_{\tilde{\nu}_L^J})
\nonumber\\&&-\sqrt{2}g_LZ_{N_L}^{1i}(Z_{N_\nu}^{I\alpha}Z_{N_\nu}^{(K+3)\theta}+Z_{N_\nu}^{I\theta}Z_{N_\nu}^{(K+3)\alpha})
\delta (M^2_{\tilde{\nu}})^{IJ}_{LR}
[(\lambda^{KJ}_{N^c}+\lambda^{JK}_{N^c})Z_{N_L}^{3i}
+\sqrt{2}g_LZ_{N_L}^{1i}\delta^{KJ}]\nonumber\\&&\times
I^0_{111}(x_{\chi_{L_i}^{0}},x_{\tilde{\nu}_L^I},x_{\tilde{N}_R^{cJ}})
+Z_{N_\nu}^{(I+3)\theta}Z_{N_\nu}^{(F+3)\alpha}\delta (M^2_{\tilde{\nu}})^{KJ}_{RR}
[(\lambda^{IJ}_{N^c}+\lambda^{JI}_{N^c})Z_{N_L}^{3i}+\sqrt{2}g_LZ_{N_L}^{1i}\delta^{IJ}]\nonumber\\&&\times
[(\lambda^{FK}_{N^c}+\lambda^{KF}_{N^c})Z_{N_L}^{3i}+\sqrt{2}g_LZ_{N_L}^{1i}\delta^{KF}]
I^0_{111}(x_{\chi_{L_i}^{0}},x_{\tilde{N}_R^{cK}},x_{\tilde{N}_R^{cJ}})\Big),
\nonumber\\&&
\delta (m_\nu)_{\alpha\theta}(\tilde{\nu}_L,\chi_i^0)=\frac{1}{2\Lambda^2}Z_{N_\nu}^{I\alpha}Z_{N_\nu}^{J\theta}(g_2Z_N^{2i}-g_1Z_N^{1i})^2m_{\chi_i^0}
\delta (M^2_{\tilde{\nu}})^{IJ}_{LL}I^0_{111}(x_{\chi_i^0},x_{\tilde{\nu}_L^I},x_{\tilde{\nu}_L^J}).\label{SNneu}
\end{eqnarray}
The virtual Higgs-charged lepton and exotic Higgs-neutrino can also give the contributions
\begin{eqnarray}
&& 
\delta (m_\nu)_{\alpha\theta}(H_d^2,e^I)=\frac{m_{e^{I}}}{\Lambda^2}Y_l^IY_l^IZ_{N_\nu}^{I\alpha}Z_{N_\nu}^{I\theta}
\delta (M^2)_{H_d^2H_d^2}I^0_{12}(x_{e^I},x_{H_d^2}),
\nonumber\\
&&\delta (m_{\nu})_{\alpha\theta}(\varphi_L^0,\bar{P}^0_{L},\chi_{N_\nu})=
\frac{m_{\chi^\beta_{\nu}}}{2\Lambda^2}\lambda^{IJ}_{N^c}\lambda^{JK}_{N^c}
Z_{N_\nu}^{(I+3)\theta}Z_{N_\nu}^{(K+3)\alpha}(Z_{N_\nu}^{(J+3)\beta})^2\nonumber\\&&\hspace{1.6cm}\times\Big(
\delta (M^2_{\varphi_L^0\varphi_L^0})I^0_{12}(x_{\chi_{\nu}^{\beta}},x_{\varphi_L^0}),
+\delta (M^2_{\bar{P}^0_{L}\bar{P}^0_{L}})I^0_{12}(x_{\chi_{\nu}^{\beta}},x_{\bar{P}^0_{L}})\Big).\label{HIGGSneu}
\end{eqnarray}
The definition of $I^0_{12}(x_1,x_2)$ is
\begin{eqnarray}
&&i\int\frac{dk^4}{(2\pi)^4}\frac{1}{k^2-m_1^2}\frac{1}{(k^2-m_2^2)^2}=\frac{1}{\Lambda^2}I^0_{12}(x_1,x_2).
\end{eqnarray}

 In the flavor basis at tree level the neutrino mass mixing matrix is
 \begin{eqnarray}
 M_{N}=\left(\begin{array}{cc}
  0&\frac{v_u}{\sqrt{2}}(Y_{\nu})^{IJ} \\
   \frac{v_u}{\sqrt{2}}(Y^{T}_{\nu})^{IJ}  & \frac{\bar{v}_L}{\sqrt{2}}(\lambda_{N^c})^{IJ}
    \end{array}\right).
\end{eqnarray}
With the rotation matrix $Z_{N_\nu}$, the masses of neutrinos are gotten by the formula
$Z_{N_\nu}^TM_NZ_{N_\nu}=diag(m_{\nu^\alpha}), ~~ \alpha=1\dots 6$.
We use the matrix $Z_{N_\nu}^T$ in the leading order of $\varsigma$, which is defined as  $\varsigma=\frac{v_u}{\bar{v}_L}(Y_{\nu})_{3\times3}.(\lambda_{N^c})^{-1}_{3\times3}$. All the elements in $\varsigma$
 are very small ($\varsigma_{IJ}\ll1$), because they are suppressed by the tiny neutrino Yukawa $Y_\nu$. It is a good
 approximation to adopt $Z_{N_\nu}^T$ in the following form\cite{munuSSMOL}
 \begin{eqnarray}
 Z_{N_\nu}^T=\left(\begin{array}{cc}
 \mathcal{S}^T&0 \\
  0 &  \mathcal{R}^T
    \end{array}\right).
    \left(\begin{array}{cc}
 1-\frac{1}{2}\varsigma^\dag\varsigma & -\varsigma^\dag\\
  \varsigma &  1-\frac{1}{2}\varsigma\varsigma^\dag
    \end{array}\right).\label{SR}
\end{eqnarray}
We use the matrices $\mathcal{S}$ and $\mathcal{R}$ defined
 in Eq.(\ref{SR}) to diagonalize $M^{seesaw}_\nu$ and $\frac{\bar{v}_L}{\sqrt{2}}\lambda_{N^c}$
  \begin{eqnarray}
 &&\mathcal{S}^TM^{seesaw}_\nu\mathcal{S}=diag(m_{\nu_1},m_{\nu_2},m_{\nu_3}),\nonumber\\&&
 \mathcal{R}^T\frac{\bar{v}_L}{\sqrt{2}}(\lambda_{N^c})^{IJ}
 \mathcal{R}=diag(m_{\nu_4},m_{\nu_5},m_{\nu_6}).
 \end{eqnarray}
In this condition, $M^{seesaw}_\nu$ is expressed as
\begin{eqnarray}
M^{seesaw}_\nu=-\frac{v^2_u}{\bar{v}_L}(Y_{\nu})_{3\times3}(\lambda_{N^c})^{-1}_{3\times3}(Y^T_{\nu})_{3\times3}.\label{treess}
\end{eqnarray}

The one loop
corrections are calculated in the mass basis at tree level, which is $( \psi_ {\nu^I_L},\psi_{N^{cI}_R} )Z_{N_\nu}$.
Here, we obtain the sum of one loop corrections from Eqs.(\ref{SLchar},\ref{SNneu},\ref{HIGGSneu})
\begin{eqnarray}
 &&\Delta (M_{N})_{\alpha\theta}=
 \delta (m_\nu)_{\alpha\theta}(\tilde{e}_L,\tilde{e}_R,\chi_{i}^{\pm})
 +\delta (m_\nu)_{\alpha\theta}(\tilde{\nu}_L,\chi_i^0)+\delta (m_\nu)_{\alpha\theta}(H_d^2,e^I)
 \nonumber\\&&\hspace{1.9cm}+\delta (m_{\nu})_{\alpha\theta}(\tilde{\nu}_L,\tilde{N}^c_R,\chi_{L_i}^{0})+\delta (m_{\nu})_{\alpha\theta}(\varphi_L^0,\bar{P}^0_{L},\chi_{N_\nu}).
\end{eqnarray}
For neutrino mass mixing matrix, to get the sum of tree level results and one loop level corrections,
we rotate $\Delta M_N$ into the flavor basis $( \psi_ {\nu^I_L},\psi_{N^{cI}_R} )$.
Therefore the sum can be expressed as
\begin{eqnarray}
&&M_N^{sum}=M_N+Z_{N_\nu}\Delta M_N Z_{N_\nu}^T\nonumber\\&&
=\left(\begin{array}{cc}
  \Delta({m_{\nu\nu}})&\frac{v_u}{\sqrt{2}}Y_{\nu}+\Delta({m_{\nu N^c}}) \\
   (\frac{v_u}{\sqrt{2}}Y_{\nu}+\Delta({m_{\nu N^c}}))^T~~~  & \frac{\bar{v}_L}{\sqrt{2}}\lambda_{N^c}+ \Delta({m_{N^cN^c}})
 \end{array}\right).\label{oneloopC}
\end{eqnarray}
Obviously, the matrix $M_N^{sum}$ in Eq.(\ref{oneloopC}) including the one loop corrections also possesses a seesaw structure.
Similar as Eq.(\ref{treess}), at one loop level we obtain the
corrected effective light neutrino mass matrix in the following form\cite{munuSSMOL}
\begin{eqnarray}
\mathcal{M}_\nu^{eff}\approx \Delta({m_{\nu\nu}})-\Big(\frac{v_uY_{\nu}}{\sqrt{2}}+\Delta({m_{\nu N^c}})\Big)
\Big( \frac{\bar{v}_L\lambda_{N^c}}{\sqrt{2}}+ \Delta({m_{N^cN^c}})\Big)^{-1}\Big(\frac{v_uY_{\nu}}{\sqrt{2}}+\Delta({m_{\nu N^c}})\Big)^T.
\end{eqnarray}

Using the " top-down " method\cite{topdown}, from the one loop corrected effective light neutrino mass matrix $\mathcal{M}_\nu^{eff}$ we
 get the Hermitian matrix
\begin{eqnarray}
{\cal H}=(\mathcal{M}_\nu^{eff})^{\dagger}\mathcal{M}_\nu^{eff}.
\end{eqnarray}
One can diagonalize the $3\times3$ matrix ${\cal H}$ to gain three eigenvalues
\begin{eqnarray}
&&m_1^2={a\over3}-{1\over3}p(\cos\phi+\sqrt{3}\sin\phi),
\nonumber\\
&&m_2^2={a\over3}-{1\over3}p(\cos\phi-\sqrt{3}\sin\phi),
\nonumber\\
&&m_3^2={a\over3}+{2\over3}p\cos\phi.\label{massQ}
\end{eqnarray}
The concrete forms of the parameters in Eq.(\ref{massQ}) are collected here
\begin{eqnarray}
&&p=\sqrt{a^2-3b}, ~~~~~\phi={1\over3}\arccos({1\over p^3}(a^3-{9\over2}ab+{27\over2}c)),
~~~~a={\rm Tr}({\cal H}),\nonumber\\
&&b={\cal H}_{11}{\cal H}_{22}+{\cal H}_{11}{\cal H}_{33}+{\cal H}_{22}{\cal H}_{33}
-{\cal H}_{12}^2 -{\cal H}_{13}^2-{\cal H}_{23}^2,~~~~c={\rm Det}({\cal H}).
\end{eqnarray}
 For the neutrino mass spectrum, there are two possibilities in  the 3-neutrino mixing case.
 The neutrino mass spectrum with normal ordering (NO) is
\begin{eqnarray}
&&m_{\nu_1}<m_{\nu_2}<m_{\nu_3}, ~~~m_{\nu_1}^2=m_1^2,\quad m_{\nu_2}^2=m_2^2,\quad m_{\nu_3}^2=m_3^2, \nonumber\\
&&\Delta m_{\odot}^2 = m_{\nu_2}^2-m_{\nu_1}^2 ={2\over \sqrt{3}}p\sin\phi>0,\nonumber\\
&&\Delta m_{A}^2 =m_{\nu_3}^2-m_{\nu_1}^2 =p(\cos\phi+{1\over\sqrt{3}}\sin\phi)>0.
\end{eqnarray}
We also write down the neutrino mass spectrum with inverted ordering (IO)
\begin{eqnarray}
&&m_{\nu_3}<m_{\nu_1}<m_{\nu_2}, ~~~m_{\nu_3}^2=m_1^2,\quad m_{\nu_1}^2=m_2^2,\quad m_{\nu_2}^2=m_3^2, \nonumber\\
&&\Delta m_{\odot}^2 = m_{\nu_2}^2-m_{\nu_1}^2 =p(\cos\phi-{1\over\sqrt{3}}\sin\phi)>0, \nonumber\\
&&\Delta m_{A}^2 = m_{\nu_3}^2-m_{\nu_2}^2 =-p(\cos\phi+{1\over\sqrt{3}}\sin\phi)<0.
\end{eqnarray}

From the mass squared matrix ${\cal H}$, one gets
the normalized eigenvectors
\begin{eqnarray}
&&\left(\begin{array}{c}\Big(U_\nu\Big)_{11}\\
\Big(U_\nu\Big)_{21}\\\Big(U_\nu\Big)_{31}
\end{array}\right)={1\over\sqrt{|X_1|^2+|Y_1|^2+|Z_1|^2}}\left(\begin{array}{c}
X_1\\Y_1\\Z_1\end{array}\right), \nonumber\\
&&\left(\begin{array}{c}\Big(U_\nu\Big)_{12}\\
\Big(U_\nu\Big)_{22}\\\Big(U_\nu\Big)_{32}
\end{array}\right)={1\over\sqrt{|X_2|^2+|Y_2|^2+|Z_2|^2}}\left(\begin{array}{c}
X_2\\Y_2\\Z_2\end{array}\right),
\nonumber\\
&&\left(\begin{array}{c}\Big(U_\nu\Big)_{13}\\
\Big(U_\nu\Big)_{23}\\\Big(U_\nu\Big)_{33}
\end{array}\right)={1\over\sqrt{|X_3|^2+|Y_3|^2+|Z_3|^2}} \left(\begin{array}{c}
X_3\\Y_3\\Z_3\end{array}\right).
\end{eqnarray}

The concrete forms of $X_I,Y_I,Z_I$ for $I=1,2,3$ are shown here
\begin{eqnarray}
&&X_1=({\cal H}_{22}-m_{{\nu_1}}^2)({\cal H}_{33}-m_{{\nu_1}}^2)-{\cal H}_{23}^2, ~~~~Y_1={\cal H}_{13}{\cal H}_{23}-{\cal H}_{12}({\cal H}_{33}-m_{{\nu_1}}^2), \nonumber\\
&&Z_1={\cal H}_{12}{\cal H}_{23}-{\cal H}_{13}({\cal H}_{22}-m_{{\nu_1}}^2),
~~~~~~~~~~X_2={\cal H}_{13}{\cal H}_{23}-{\cal H}_{12}\Big({\cal H}_{33}-m_{{\nu_2}}^2\Big),
\nonumber\\
&&Y_2=({\cal H}_{11}-m_{{\nu_2}}^2)({\cal H}_{33}-m_{{\nu_2}}^2)-{\cal H}_{13}^2,
~~~~Z_2={\cal H}_{12}{\cal H}_{13}-{\cal H}_{23}\Big({\cal H}_{11}-m_{{\nu_2}}^2\Big),
\nonumber\\
&&X_3={\cal H}_{12}{\cal H}_{23}-{\cal H}_{13}\Big({\cal H}_{22}-m_{{\nu_3}}^2\Big),
~~~~~~~~~Y_3={\cal H}_{12}{\cal H}_{13}-{\cal H}_{23}\Big({\cal H}_{11}-m_{{\nu_3}}^2\Big),
\nonumber\\
&&Z_3=({\cal H}_{11}-m_{{\nu_3}}^2)({\cal H}_{22}-m_{{\nu_3}}^2)-{\cal H}_{12}^2.
\end{eqnarray}
The mixing angles among three tiny neutrinos can be defined as follows
\begin{eqnarray}
&&\sin\theta_{13}=\Big|\Big(U_\nu\Big)_{13}\Big|,~~~~~~~~~~~~~~~\cos\theta_{13}=\sqrt{1-\Big|\Big(U_\nu\Big)_{13}\Big|^2},\nonumber\\
&&
\sin\theta_{23}={\Big|\Big(U_\nu\Big)_{23}\Big|\over\sqrt{1-\Big|\Big(U_\nu\Big)_{13}\Big|^2}},~~~~~~
\cos\theta_{23}={\Big|\Big(U_\nu\Big)_{33}\Big|\over\sqrt{1-\Big|\Big(U_\nu\Big)_{13}\Big|^2}},\nonumber\\
&& \sin\theta_{12}={\Big|\Big(U_\nu\Big)_{12}\Big|\over\sqrt{1-\Big|\Big(U_\nu\Big)_{13}\Big|^2}},~~~~~~
\cos\theta_{12}={\Big|\Big(U_\nu\Big)_{11}\Big|\over\sqrt{1-\Big|\Big(U_\nu\Big)_{13}\Big|^2}}.
\end{eqnarray}

\section{numerical results}
In this section, we discuss the numerical results for the neutrinos including three mixing angles
and two mass squared differences.
Using BLMSSM, we have studied several processes in our precious works,
such as the lightest neutral CP-even Higgs mass
and the charged lepton flavor violating processes.
Because the masses of light neutrinos are very tiny at $10^{-1}$ TeV order, the used parameters should be precise enough.
In this work, the tiny neutrino Yukawa $Y_\nu$ can give contributions to light neutrino masses at tree level through
the see-saw mechanism. Therefore, $Y_\nu$ are important parameters and should be considered earnestly.

We show the used parameters
\begin{eqnarray}
&&B_H = 800{\rm GeV},~~~
\mu_H = 650{\rm GeV},~~~
m_1 = -1{\rm TeV},~~~
m_2 = -500{\rm GeV},~~~
\lambda_{N^c} = 1,\nonumber\\&&
(m_{\tilde{L}}^2)_{11} = 1347.203^2 {\rm GeV^2},~~
(m_{\tilde{L}}^2)_{22} = 1238.028^2{\rm GeV^2},~~
(m_{\tilde{L}}^2)_{33} = 1464.3458^2{\rm GeV^2},
\nonumber\\&&
(m_{\tilde{R}}^2) = \delta_{ij}{\rm TeV^2},~~~
(A_l) = -3\delta_{ij}{\rm TeV},~~~
(A'_l)=3\delta_{ij}{\rm TeV},~~~
A_{N^c} = A_N = 2\delta_{ij}{\rm TeV},
\nonumber\\&&
m_{\tilde{e}_L^1}= 3250{\rm GeV},~~~
m_{\tilde{e}_L^2}= 3250{\rm GeV},~~~
m_{\tilde{e}_L^3}= 3260{\rm GeV},~~~
m_{\tilde{e}_R^1} = 3600{\rm GeV},\nonumber\\&&
m_{\tilde{e}_R^2} = 3400{\rm GeV},~~~
m_{\tilde{e}_R^3} = 3200{\rm GeV},~~~
m_{\tilde{\nu}_L^i}=m_{\tilde{\nu}_R^i} = 2{\rm TeV},~~~
m_{H_d^2} = 1{\rm TeV},\nonumber\\&&
B_L = 1{\rm TeV},~~~
\tan\beta =10,~~~
\tan\beta_L = 2,~~~
v_{L_t} = 3{\rm TeV},~~~(m_{\tilde{N}}^2) =  \delta_{ij}{\rm TeV}^2
,\nonumber\\&&
\mu_L = 3{\rm TeV},~~~
m_L = 2{\rm TeV},~~~
m_{\varphi_L^0} = 1{\rm TeV},~~~
m_{\bar{P}_L^0} = 1{\rm TeV},~~~g_L = 1/6.
\end{eqnarray}

\subsection{NO spectrum}
At first we study the NO spectrum with the supposition for neutrino Yukawa couplings
\begin{eqnarray}
&&(Y_\nu)^{11} = 1.295656092\times10^{-6},~~~
(Y_\nu)^{22} = 1.595819186\times10^{-6},\nonumber\\&&
(Y_\nu)^{33} = 1.696349655\times10^{-6},~~~
(Y_\nu)^{12}=  9.774376457\times10^{-8},\nonumber\\&&
(Y_\nu)^{13} = 6.418350381\times10^{-8},~~~
(Y_\nu)^{23} = 4.056255181\times10^{-8},
\end{eqnarray}
and issue the numerical results for light neutrino masses and mixing
angles
\begin{eqnarray}
&&\hspace{1.0cm}	
|\Delta m_{A}^2| =2.4707\times10^{-3}{\rm eV^2},~~~~~~~~~~~\Delta m_{\odot}^2 =7.5344\times10^{-5}{\rm eV^2}, \nonumber\\
&&\hspace{0.8cm}\sin^2\theta_{12}=0.3221,~~~~~~~~~ 	\sin^2\theta_{13}=0.0247,~~~~~~~~~	
\sin^2\theta_{23}=0.5240, \nonumber\\&& m_{\nu_1}=1.9503\times10^{-1}{\rm eV},~~ m_{\nu_2}=1.9522\times10^{-1}{\rm eV},~~
 m_{\nu_3}=2.0126\times10^{-1}{\rm eV}.
\end{eqnarray}
The diagrams are plotted near this point that satisfies the experiment constraints for light neutrinos.

\begin{figure}[h]
\setlength{\unitlength}{1mm}
\centering
\includegraphics[width=2.9in]{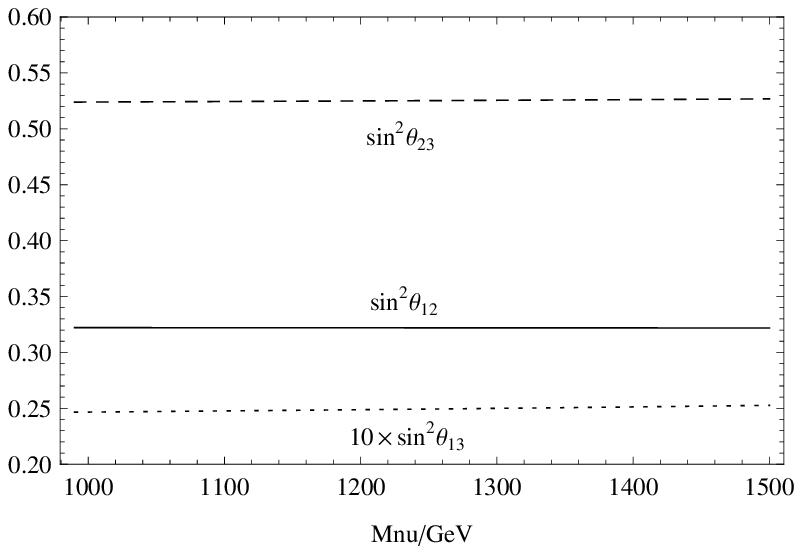}~~~ \includegraphics[width=2.8in]{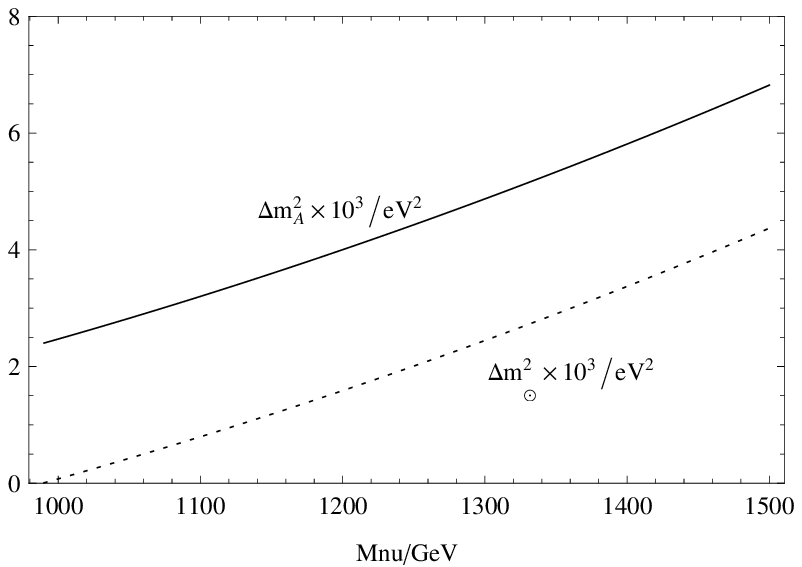}
\caption[]{With NO assumption for neutrino mass spectrum, we plot the neutrino mixing angles
and mass squared differences versus $Mnu$. In the left diagram
$\sin^2\theta_{12}$(solid line), $\sin^2\theta_{13}$(dotted line) and $\sin^2\theta_{23}$(dashed line)
 versus $Mnu$, and in the right diagram $\Delta m^2_A$(solid line) and $\Delta m^2_\odot$(dotted line)
versus $Mnu$, respectively.}\label{NOMnu}
\end{figure}
The parameters $(m_{\tilde{N}}^2)$ are relevant to the masses of right-handed scalar neutrinos which
give contributions to neutrino mass matrix through the coupling of neutrino-sneutrino-lepton neutralino.
 Using the assumption $(m_{\tilde{N}}^2)=Mnu^2\delta_{ij}$,
 we discuss how the scalar neutrinos affect the results from
 $(m_{\tilde{N}}^2)$ in the Fig.\ref{NOMnu}. In the left diagram of Fig.\ref{NOMnu},
 the neutrino mixing angles $\sin^2\theta_{12}$, $10\times\sin^2\theta_{13}$ and $\sin^2\theta_{23}$ are represented by the
solid line, dotted line and dashed line respectively. These three lines all vary weakly versus $Mnu$ in the region $1000\sim1500$ {\rm GeV}.
The values of $\sin^2\theta_{12}$ and $\sin^2\theta_{23}$ are around 0.32 and 0.52 respectively.
$\sin^2\theta_{13}$ is the smallest one and near 0.025. In this region of $Mnu$, the three mixing angles are all satisfy the experiment bounds.
We show the theoretical predictions for $\Delta m^2_A$ and $\Delta m^2_\odot$ versus $Mnu$ by the solid line
 and dotted line in the right diagram of Fig.\ref{NOMnu}. They are both obviously increasing
 functions of the $Mnu$ from 1000 GeV to 1500 GeV. Considering the experiment
bounds of $\Delta m^2_A$ and $\Delta m^2_\odot$, the applicable range of $Mnu$ is near 1000 GeV.

\begin{figure}[h]
\setlength{\unitlength}{1mm}
\centering
\includegraphics[width=2.9in]{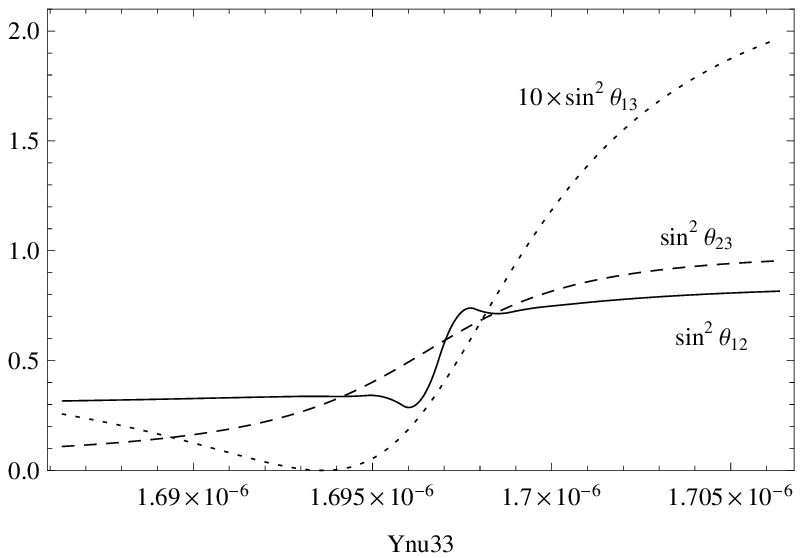}~~~ \includegraphics[width=2.8in]{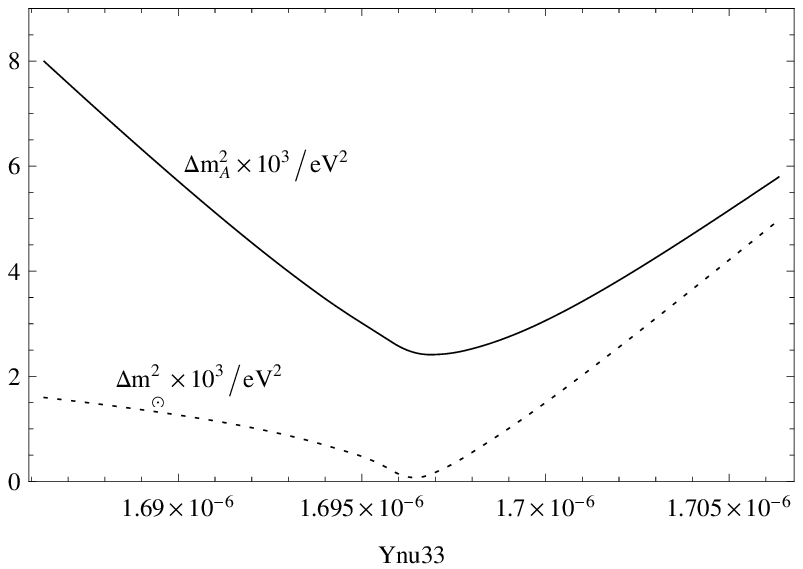}
\caption[]{With NO assumption for neutrino mass spectrum, we plot the neutrino mixing angles
and mass squared differences versus $Ynu33$. In the left diagram
$\sin^2\theta_{12}$(solid line), $\sin^2\theta_{13}$(dotted line) and $\sin^2\theta_{23}$(dashed line)
 versus $Ynu33$, and in the right diagram $\Delta m^2_A$(solid line) and $\Delta m^2_\odot$(dotted line)
versus $Ynu33$, respectively.}\label{NOYnu33}
\end{figure}
Though the neutrino Yukawa couplings are tiny, they can give important contributions to the neutrino mixing angles
and masses, because they contribute at tree level. Here we discuss the effects from $(Y_\nu)^{33}$ with the definition
$(Y_\nu)^{33}=Ynu33$. When $Ynu33$ varies from $1.686\times10^{-6}$ to $1.706\times10^{-6}$, the behaviors
of $\sin^2\theta_{12}$, $\sin^2\theta_{13}$ and $\sin^2\theta_{23}$ are studied numerically in the left diagram of Fig.\ref{NOYnu33}.
The solid line represents $\sin^2\theta_{12}$, and changes softly except the region $(1.695\sim1.699)\times10^{-6}$.
The values of $\sin^2\theta_{23}$ denoted by the dashed line are the increasing functions of $Ynu33$, and vary from
0.1 to 0.95. The dotted line represents $10\times\sin^2\theta_{13}$, which is very tiny near the point $Ynu33=1.694\times10^{-6}$.
For the three mixing angles, the suitable region of $Ynu33$ is near $1.696\times10^{-6}$. The right diagram in Fig.\ref{NOYnu33}
shows the behaviors of $\Delta m^2_A$(solid line) and $\Delta m^2_\odot$(dotted line) versus $Ynu33$.
The values of $\Delta m^2_A$ and $\Delta m^2_\odot$ arrive at small numerical results as $Ynu33$ between $1.696\times10^{-6}$ and $1.697\times10^{-6}$.

\begin{figure}[h]
\setlength{\unitlength}{1mm}
\centering
\includegraphics[width=2.9in]{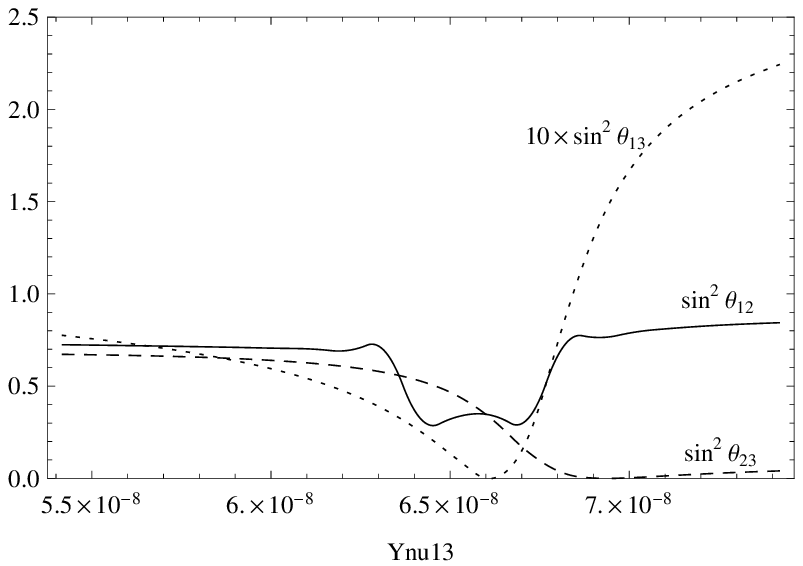}~~~ \includegraphics[width=2.8in]{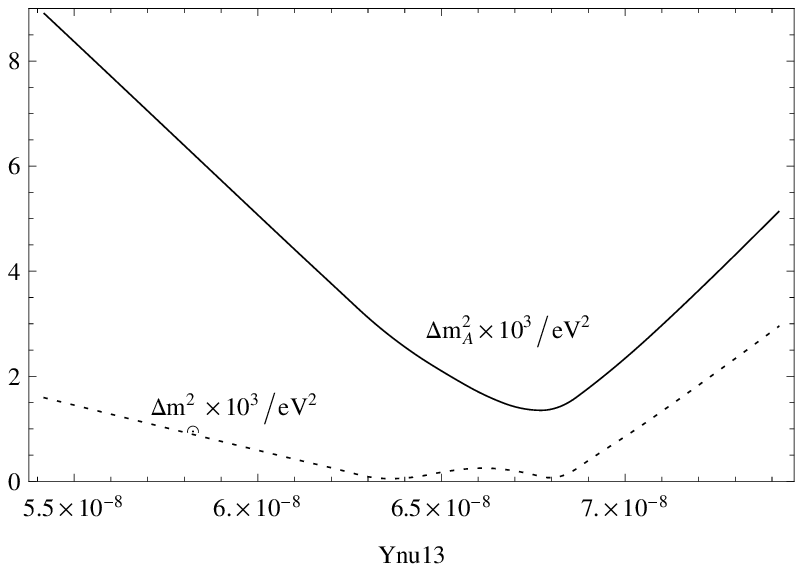}
\caption[]{With NO assumption for neutrino mass spectrum, we plot the neutrino mixing angles
and mass squared differences versus $Ynu13$. In the left diagram
$\sin^2\theta_{12}$(solid line), $\sin^2\theta_{13}$(dotted line) and $\sin^2\theta_{23}$(dashed line)
 versus $Ynu13$, and in the right diagram $\Delta m^2_A$(solid line) and $\Delta m^2_\odot$(dotted line)
versus $Ynu13$, respectively.}\label{NOYnu13}
\end{figure}
Besides the diagonal elements of $Y_\nu$, the non-diagonal elements of $Y_\nu$ are also important parameters. Here, we discuss how $(Y_\nu)^{13}=Ynu13$
influences the theoretical predictions on the neutrino mixing angles and mass squared differences in the Fig.\ref{NOYnu13}. In the left diagram,
the solid line ($\sin^2\theta_{12}$) looks like "U" in the $Ynu13$ region $(6.2\times10^{-8}\sim6.9\times10^{-8})$, and in the other regions
the values of $\sin^2\theta_{12}$ are about 0.75 and relatively stable. The dashed line denoting $\sin^2\theta_{23}$ is the decreasing function of the increasing $Ynu13$ during the region $(5.4\times10^{-8} \sim 7.4\times10^{-8})$. We show the values for $10\times\sin^2\theta_{13}$ by the dotted line which varies from almost zero to 2.25.
Near the point $Ynu13=6.6\times10^{-8}$, the values of $10\times\sin^2\theta_{13}$ are very small.
In the right diagram of Fig.\ref{NOYnu13}, the dotted line($\Delta m^2_\odot$) is small
in the range  $6.4\times10^{-8}<Ynu13<6.9\times10^{-8}$. The values of $\Delta m^2_A\times10^3/{\rm eV}^2$
represented by the solid line vary from 2.5 to 8.0. Taking into account the
 neutrino experiment bounds, the appropriate $Ynu13$ value is around $6.4\times10^{-8}$.

\subsection{IO spectrum}
With the neutrino mass spectrum being IO, the neutrino mass
squared differences and mixing angles are also studied numerically here. Using the following parameters
\begin{eqnarray}
&&
(Y_\nu)^{11} = 1.306444732\times10^{-6},~~~
(Y_\nu)^{22} = 1.594718138\times10^{-6},\nonumber\\&&
(Y_\nu)^{33} = 1.694606167\times10^{-6},~~~
(Y_\nu)^{12}=  9.556253970\times10^{-8},\nonumber\\&&
(Y_\nu)^{13} = 5.940807045\times10^{-8},~~~
(Y_\nu)^{23} = 3.061881997\times10^{-8},
\end{eqnarray}
we get the numerical results for the  neutrino masses and mixing angles at this point. The obtained numerical
results are shown in the following form
\begin{eqnarray}
&&\sin^2\theta_{12}=0.2806,~~~ 	\sin^2\theta_{13}=0.0212,~~~	
\sin^2\theta_{23}=0.4939, \nonumber\\&&	
|\Delta m_{A}^2| =2.5819\times10^{-3}{\rm eV^2},~~~~~~\Delta m_{\odot}^2 =7.6694\times10^{-5}{\rm eV^2}, \nonumber\\&& m_{\nu_1}=1.9510\times10^{-1}{\rm eV},~~~~~~~~~~~~	 m_{\nu_2}=1.9529\times10^{-1}{\rm eV},\nonumber\\&&
 m_{\nu_3}=1.8857\times10^{-1}{\rm eV}.
\end{eqnarray}
The light neutrino masses are all at the order of $10^{-1}$ eV.

\begin{figure}[h]
\setlength{\unitlength}{1mm}
\centering
\includegraphics[width=2.9in]{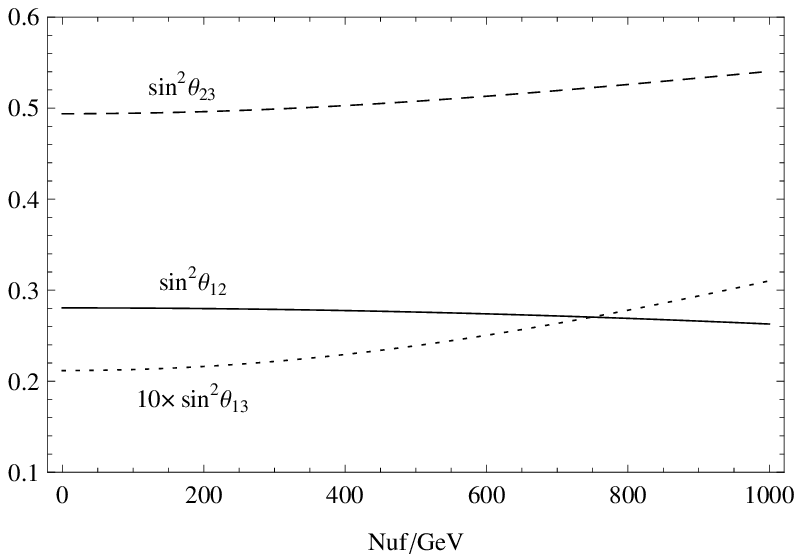}~~~ \includegraphics[width=2.8in]{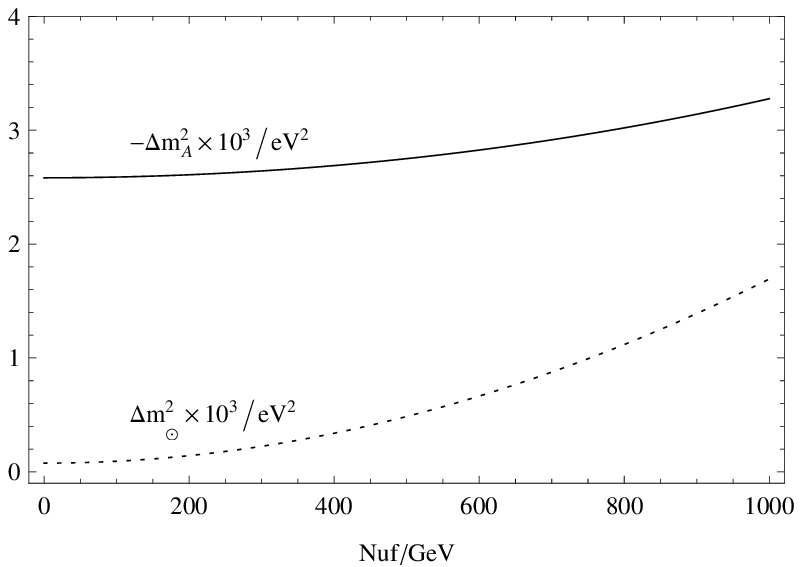}
\caption[]{With IO assumption for neutrino mass spectrum, we plot the neutrino mixing angles
and mass squared differences versus $Nuf$. In the left diagram
$\sin^2\theta_{12}$(solid line), $\sin^2\theta_{13}$(dotted line) and $\sin^2\theta_{23}$(dashed line)
 versus $Nuf$, and in the right diagram $-\Delta m^2_A$(solid line) and $\Delta m^2_\odot$(dotted line)
versus $Nuf$, respectively.}\label{IONuf}
\end{figure}
Here, we discuss the effects from the non-diagonal elements of $(m_{\tilde{N}}^2)$, and suppose
$(m_{\tilde{N}}^2)_{ij}=Nuf^2,~ for~ i,j=1,2,3~ and~ i\neq j$.
In the left diagram of Fig.\ref{IONuf}, we depict the solid line($\sin^2\theta_{12}$), dotted line ($10\times\sin^2\theta_{13}$) and dashed line($\sin^2\theta_{23}$), respectively. These three lines change mildly with $Nuf$ varying from 0 to 1000 GeV.
For the parameter $Nuf$, the dotted line and dashed line
are both increasing functions, and the solid line is the decreasing function. The neutrino mass
squared differences versus $Nuf$ denoted by the solid line($-\Delta m^2_A$) and dotted line($\Delta m^2_\odot$) are plotted
in the right diagram of Fig.\ref{IONuf}. $\Delta m^2_\odot$ increases a little faster than $-\Delta m^2_A$ with the increasing $Nuf$.
In this parameter space, $Nuf$ should be no more than 200GeV as shown in the Fig.\ref{IONuf}.

\begin{figure}[h]
\setlength{\unitlength}{1mm}
\centering
\includegraphics[width=2.9in]{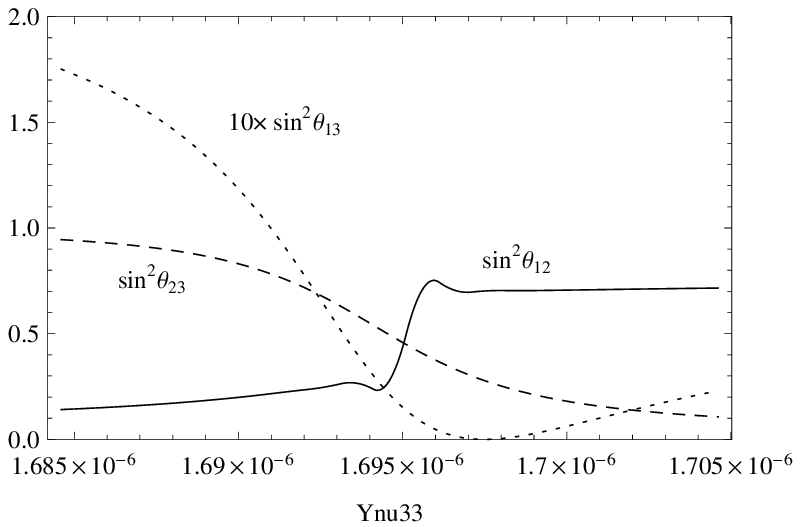}~~~ \includegraphics[width=2.8in]{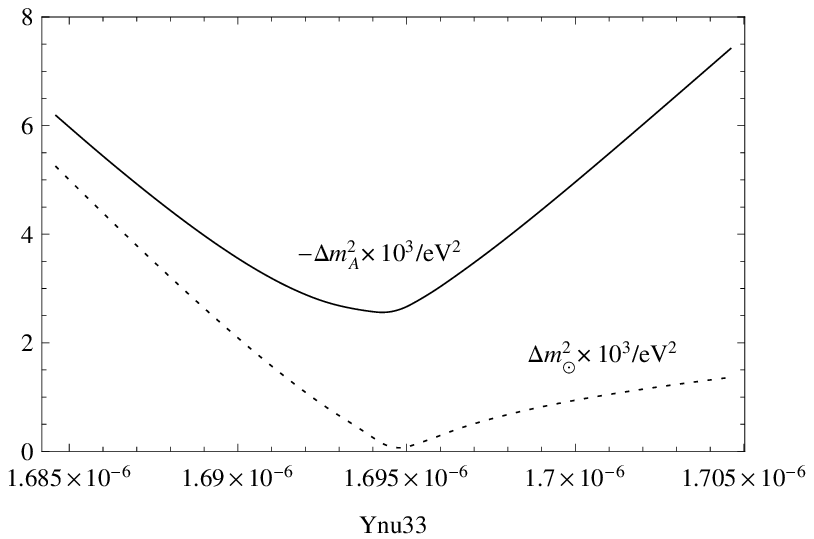}
\caption[]{With IO assumption for neutrino mass spectrum, we plot the neutrino mixing angles
and mass squared differences versus $Ynu33$. In the left diagram
$\sin^2\theta_{12}$(solid line), $\sin^2\theta_{13}$(dotted line) and $\sin^2\theta_{23}$(dashed line)
 versus $Ynu33$, and in the right diagram $-\Delta m^2_A$(solid line) and $\Delta m^2_\odot$(dotted line)
versus $Ynu33$, respectively.}\label{IOYnu33}
\end{figure}
\begin{figure}[h]
\setlength{\unitlength}{1mm}
\centering
\includegraphics[width=2.9in]{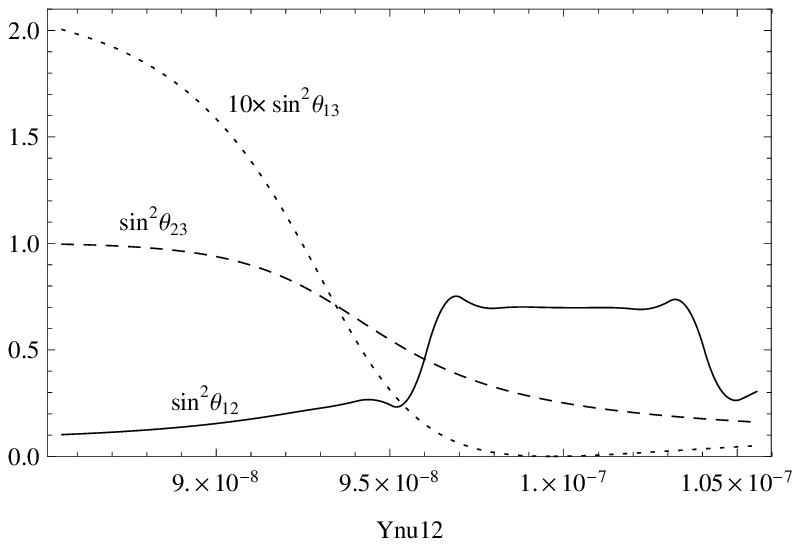}~~~ \includegraphics[width=2.8in]{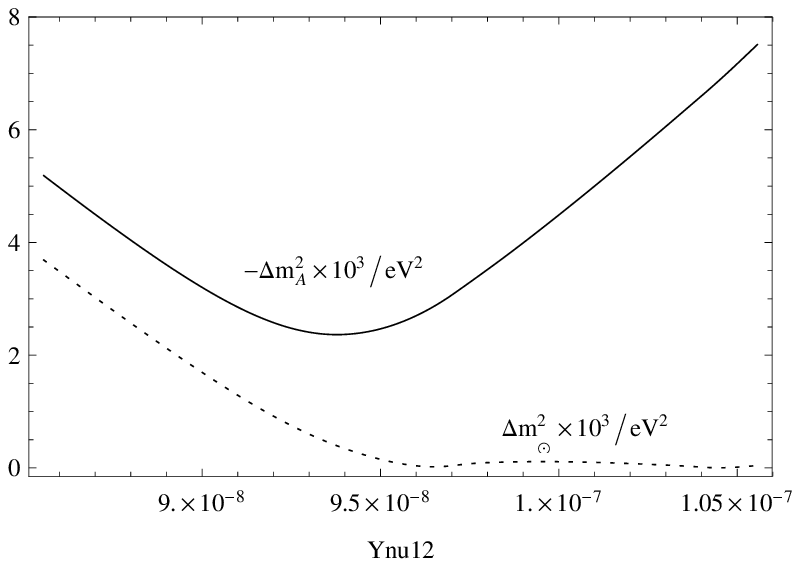}
\caption[]{With IO assumption for neutrino mass spectrum, we plot the neutrino mixing angles
and mass squared differences versus $Ynu12$. In the left diagram
$\sin^2\theta_{12}$(solid line), $\sin^2\theta_{13}$(dotted line) and $\sin^2\theta_{23}$(dashed line)
 versus $Ynu12$, and in the right diagram $-\Delta m^2_A$(solid line) and $\Delta m^2_\odot$(dotted line)
versus $Ynu12$, respectively.}\label{IOYnu12}
\end{figure}

Here, we also discuss how the diagonal element $(Y_{\nu})^{33}=Ynu33$ influences the theoretical
predictions on the neutrino mixing angles and mass squared differences in the Fig.\ref{IOYnu33}.
From the solid line($\sin^2\theta_{12}$), dotted line($10\times\sin^2\theta_{13}$) and dashed line($\sin^2\theta_{23}$) in the left diagram,
we should take $Ynu33$ around $1.695\times10^{-6}$.
In the right diagram, both the solid line($-\Delta m^2_A$) and
the dotted line($\Delta m^2_\odot$) reach small values near the point $Ynu33=1.695\times10^{-6}$.
The non-diagonal element $(Y_{\nu})^{12}=Ynu12$ can obviously influence the numerical results for the neutrinos.
The mixing angles($\sin^2\theta_{12}, 10\times\sin^2\theta_{13},\sin^2\theta_{23}$) versus $Ynu12$ are plotted by the
solid, dotted and dashed line in the left diagram of Fig.\ref{IOYnu12}.
We show the neutrino mass squared differences $-\Delta m^2_A$(the solid line) and $\Delta m^2_\odot$(the dashed line)
in the right diagram of Fig.\ref{IOYnu12}. From the both diagrams and neutrino experiment bounds,
the appropriate value of $Ynu12$ should be around $9.6\times10^{-8}$. As it is well known that the light neutrino masses are very tiny and there are
five experiment constraints(three mixing angles and two mass squared differences), the obtained suitable parameter space is narrow.

\section{summary}
The neutrino experiment data from
both solar and atmospheric neutrino experiments show that neutrinos have tiny masses and three mixing
angles including two large mixing angles and one small mixing angle. The SM can not solve the neutrino experiment data, and
physicists consider SM should be the low energy effective theory of a large model. So, the SM should be extended.
One of the supersymmetric extensions of the SM is BLMSSM which has local gauged $B$ and $L$ symmetries. In this model, we have
studied some processes in our previous works\cite{BLMSSM2,BLBarZee,BLMSSM3,BL750}. In this work,
with the mass insertion approximation the one loop corrections to the
neutrino mixing matrix are researched.

In the BLMSSM, the tree level neutrino mass mixing matrix is obtained in our previous work\cite{BLCB}.
The obtained one loop corrections include:
1. the virtual slepton-chargino corrections; 2. the virtual sneutrino-lepton neutralino corrections;
3. the virtual sneutrino-neutralino corrections;
4. the virtual Higgs-charged lepton corrections; 5. the exotic Higgs-neutrino corrections. 
We get the sum of the tree and one loop contributions to the neutrino mixing matrix.
The one loop corrected effective light neutrino mass matrix $\mathcal{M}_\nu^{eff}$ is deduced.
Using the ¡°top-down¡± method, we give the formulae for the neutrino masses and mixing angles.
For neutrino mass spectrum, both NO and IO conditions are discussed numerically. In our used
parameter space, the obtained numerical results for the neutrino three mixing angles and two mass squared differences
can account for the corresponding experiment data. Our results imply that the light neutrino masses are at the order of
$10^{-1}$ eV.

{\bf Acknowledgements}

   This work has been supported by the Major Project of NNSFC(NO.11535002) and NNSFC(NO.11275036),
   the Open Project Program of State Key Laboratory of Theoretical Physics, Institute of Theoretical Physics, Chinese
Academy of Sciences, China (No.Y5KF131CJ1), the Natural Science Foundation
of Hebei province with Grant No. A2013201277 and No. A2016201010 and the Found of Hebei province with
the Grant NO. BR2-201 and the
Natural Science Fund of Hebei University with Grants
No. 2011JQ05 and No. 2012-242, Hebei Key Lab of Optic-Electronic Information and Materials,
 the midwest universities comprehensive strength promotion project.

\end{document}